\documentclass{osa-article}

\journal{osajournal}

\articletype{Research Article}
\usepackage{lineno}
\usepackage{wrapfig}
\usepackage{xspace}
\usepackage{subfigure}
\usepackage{graphicx}
\usepackage{multirow}
\usepackage{amsmath}
\usepackage{rotating}

\newcommand\adeg{\mbox{$^\circ$}\xspace}% 

\begin{document}
\title{High Efficiency Echelle Gratings for the Far Ultraviolet}

\author{Nicholas Kruczek\authormark{1,*}, Drew M. Miles\authormark{2,3}, Brian Fleming\authormark{1}, Randall McEntaffer\authormark{2,3}, Kevin France\authormark{1,4}, Fabien Gris\'{e}\authormark{2}, Stephan McCandliss\authormark{5}}

\address{\authormark{1}Laboratory for Atmospheric and Space Physics, University of Colorado, 600 UCB, Boulder, CO 80309, USA\\
\authormark{2}Pennsylvania State University, University Park, PA, USA\\
\authormark{3}RFD Optics, LLC, University Park, PA, USA\\
\authormark{4}Center for Astrophysics and Space Astronomy, Boulder, CO, USA \\
\authormark{5}Department of Physics and Astronomy, The Johns Hopkins University, Baltimore, MD, USA}

\email{\authormark{*}nicholas.kruczek@colorado.edu}

\begin{abstract}
Modern grating manufacturing techniques suffer from inherent issues that limit their peak efficiencies. The anisotropic etching of silicon facilitates the creation of custom gratings that have sharp and atomically smooth facets, directly addressing these issues. We describe work to fabricate and characterize etched silicon echelles optimized for the far ultraviolet (FUV; 90 – 180 nm) bandpass. We fabricate two echelles that have similar parameters to the mechanically ruled grating flown on the CHESS sounding rocket. We demonstrate a 42\% increase in peak order efficiency and an 83\% decrease in interorder scatter using these gratings. We also present analysis on where the remaining efficiency resides. These demonstrated FUV echelle improvements benefit the faint source sensitivity and high-resolution performance of future UV observatories.
\end{abstract}

\section{Introduction} \label{intro}
Echelles used in the far ultraviolet (FUV; 90 – 180 nm) bandpass historically have been fabricated using mechanical ruling techniques. This process uses a diamond stylus drawn across the face of a substrate to cut and trace the grooves, producing sharp angled facets with high diffractive efficiency for select wavelengths. Traditional high-resolution FUV spectrographs (e.g., $HST$-STIS; \cite{Woodgate98}) use mechanically ruled echelles, but instrumental efficiencies are low due to scatter and loss of light to off-blaze orders. Holographic ruling is not suited for these high resolution applications due its inability to rule the required low groove densities. 

There are several sources of error that contribute to the efficiency losses of mechanically ruled gratings. The two sources of interest to this work arise from stylus wear and groove non-uniformities. Stylus wear, which results in a slightly different groove shape at the beginning than at the end of the set and rule process, is particularly problematic for large-area gratings. Groove non-uniformities can be periodic, from the periodic mechanisms that drive the ruling engine, or random, from surface and tip irregularities or from variations in the ruling conditions~\cite{Dunning80,Mi17,Gao21}. The resulting rounded edges and imprecise groove locations shift power into adjacent orders, scattered light, and diffraction ghosts, reducing the instrument throughput~\cite{Loewen95,Palmer14}. For example, the echelle modes of $HST$-STIS scatter up to 20$\%$ of the power in each order across $\gtrsim$ 400 pixels in the cross-dispersion direction, which, when compounded over tens of orders with mixed scattered emission and absorption lines, amounts to prohibitive, non-uniform backgrounds for observing faint sources \cite{Content96,stisbad}. 

Current work on X-ray reflection gratings leverages the inherent crystalline structure of silicon to produce X-ray gratings with diffraction efficiencies within 90$\%$ of the theoretical maximum~\cite{Miles18}. This same process can be extended to the FUV, where groove efficiencies on the order of 80\% have been demonstrated for low-order gratings~\cite{Sheng09}. The technique produces gratings with atomically-smooth facets, resulting in more light diffracted into the blaze order and reduced scatter relative to state-of-the-art ruling methods. The technique relies on a free-form electron-beam (e-beam) patterning process that enables the creation of custom groove traces. This pattern is then transferred into Si using an anisotropic KOH etch, producing sharp and atomically smooth grating facets at custom blaze angles. A description of the recording process is presented in Section~\ref{fab_deets}. For additional details see, e.g., Refs.~\cite{Miles18}, ~\cite{Franke97}, and references therein. 

To demonstrate the gains provided by this technique, we fabricated two new versions of the echelle flown on the Colorado High-resolution Echelle Stellar Spectrograph (CHESS) sounding rocket payload~\cite{France16}. The payload was designed to operate from 100 -- 160 nm. CHESS used a replica of a mechanically ruled grating, fabricated by Richardson Gratings~\cite{Kruczek17, Kruczek18}. This optic had a groove density of 89 grooves mm$^{-1}$, an angle of incidence ($\alpha$) of 63$^{\circ}$, and a ruled area of 102 $\times$ 102 mm. The optic was operated in Littrow configuration with a tilt ($\gamma$) of $\sim$6$^{\circ}$ to redirect the incoming beam to the cross dispersing grating. As a part of the CHESS flight program a spare gold coated echelle was ordered and this is the optic we use for comparison to the new etched gratings.

The layout of the paper is as follows: Section~\ref{fab_deets} describes the fabrication process for the echelles studied in this work. Section~\ref{CHESS} provides details on the efficiency measurement procedure, with efficiency results presented in Section~\ref{chess_eff}. Section~\ref{diffpatt} describes our efforts to characterize the full diffraction arc produced by these new gratings. We discuss aperture diffraction effects observed in the diffracted spot in Section~\ref{sptstr}. We then compare the CHESS in-instrument performance of an etched echelle to that of the mechanically-ruled flight echelle in Section~\ref{persec}, where we also demonstrate the gains provided by this new technology to a future large UV/O/IR mission. We summarize our results and discuss future work, including plans for the fabrication of gratings with holographic solutions, in Section~\ref{conc}.

\section{Echelle Fabrication} \label{fab_deets}
\begin{figure}[t]
\includegraphics[width=\textwidth]{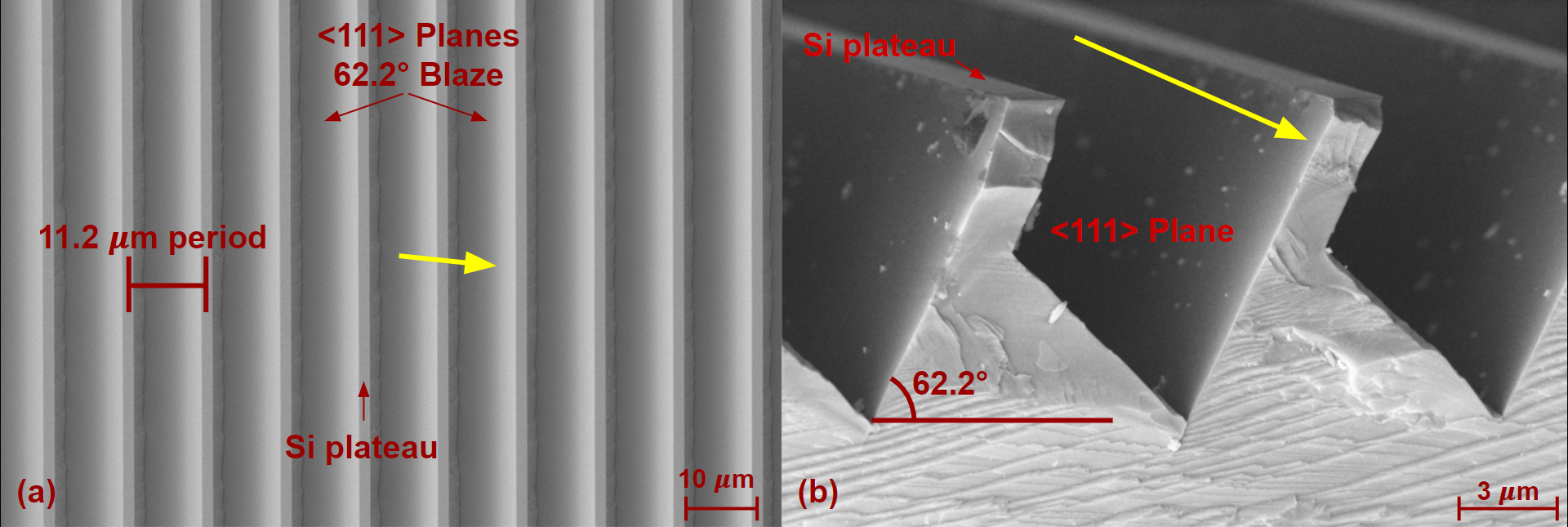}
\caption{Scanning electron microscope images of an example KOH-etched echelle. The yellow arrows in both figures show the vector of the incoming beam. (a) Top view, labeled for reference are the period, blazed facets, and an Si plateau. (b) Side view, showing the groove profile with the plateau, illuminated facet, and blaze angle labeled. Obtaining the side view images requires that the grating gets cleaved, so the grating shown in these images is a test echelle. The specks on the facets in the image are KOH crystals that are removed from the grating surface prior to coating. }
  \label{etchex}
\end{figure}
Fabrication begins with the acquisition of crystalline Si substrates with the appropriate crystal orientation; the \{111\} planes of crystalline Si define the grating facet angle because they etch at a significantly slower rate than any other Si plane. Substrates can therefore be cut from a Si boule with the \{111\} planes oriented to match the desired grating blaze angle. The procured substrates are then coated with a layer of Si$_3$N$_4$, which serves as a hard mask for the KOH etch process, and a resist that provides a reactive layer for the e-beam. The coated substrate is then installed in the e-beam writer and the writer is programmed to expose the traces of the grooves. After e-beam exposure, the substrate is developed so that either the resist that was exposed to the e-beam is removed (for a positive-tone resist) or remains on the substrate surface (for a negative-tone resist). In both cases, the final result is a grating pattern defined by the remaining resist and the exposed nitride. The resulting pattern is then dry-etched into the nitride itself, forming the hard mask, and any residual resist is stripped. Finally, the substrate is etched in the bath of KOH, forming the blazed facets defined by the \{111\} planes for the grooves traced in the nitride hard mask. 

Of particular importance to this process is that some residual nitride is necessary to define the grating groove pattern after the dry etch, otherwise the entire Si surface would be etched when exposed to KOH. While the nitride is ultimately removed in a process step after KOH exposure, narrow ($\sim$5-10\% of the period), flat Si plateaus remain at the top of the groove facets, representing the areas that were masked by the nitride. These plateaus do not contribute to the blazed diffraction pattern and set an upper limit on groove efficiency; fabrication optimization often focuses on minimizing the width of these plateaus to reduce their impact on grating performance. As a demonstration of these features, top-down and side views of an etched grating are shown in Fig.~\ref{etchex}. 

\begin{figure}[b]
\centering
\subfigure{\includegraphics[width=0.5\textwidth]{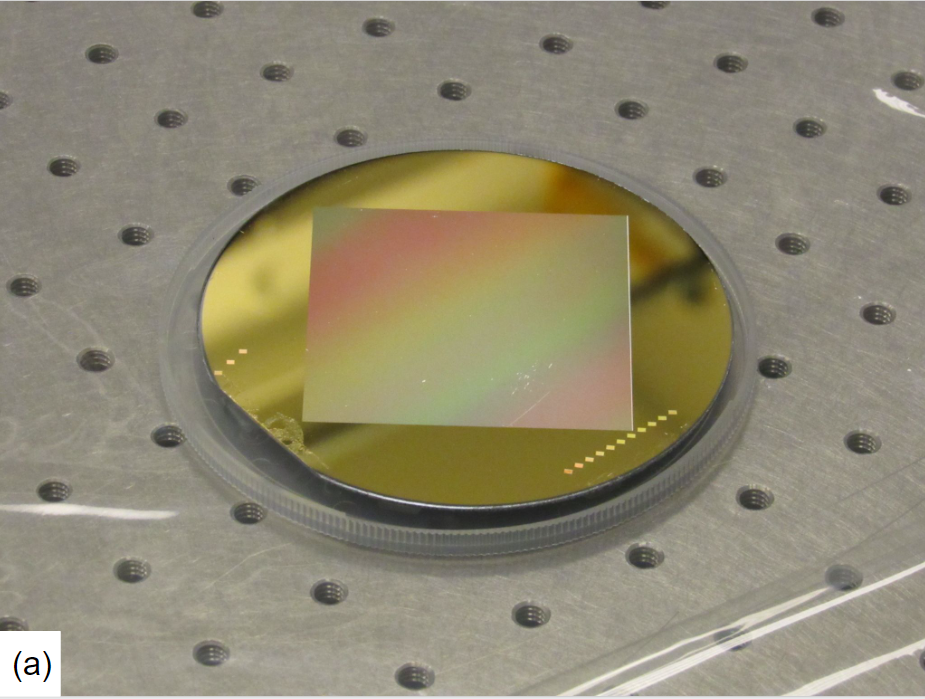}}
\subfigure{\includegraphics[width=0.481\textwidth]{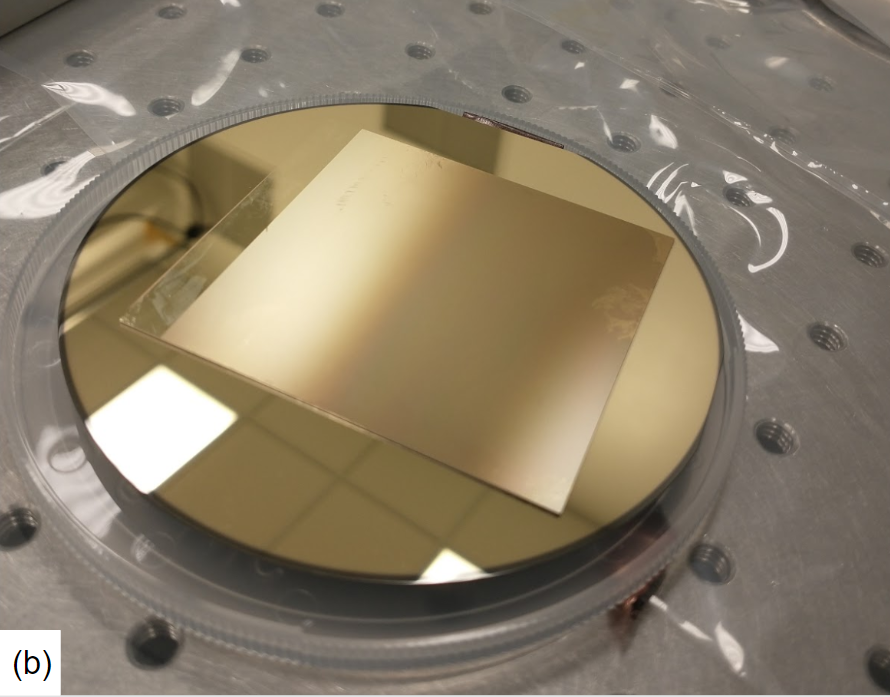}}
\caption{The gold coated V1 (a; 1.5 $\mu$m wide Si plateaus) and V2 (b; 0.4 $\mu$m wide Si plateaus) echelle gratings fabricated by RFD Optics, LLC.}
\label{echelles}
\end{figure}

Two echelles were fabricated by RFD Optics, LLC for this work (Fig.~\ref{echelles}). Both have a 60 $\times$ 60 mm ruled area, which was selected to optimize fabrication time to fit within the project schedule. They were etched from 100 mm diameter by 1 mm thick Si wafers oriented such that the \{510\} crystal plane faced upwards to set the \{111\} plane at 62.2 $\pm$ 1.0$^{\circ}$. This degree level error on the blaze angle is set by the dicing accuracy of the vendor. This was less than a degree off from the CHESS blaze angle while having the benefit of being an off the shelf wafer, saving cost and lead time on procurement. Both gratings fabricated in this work were made from phosphorus-doped Cz substrates with a resistivity $>$1 ohm-cm. The wafers were polished by the vendor to have a surface roughness of 1 nm and a bow $<$40 $\mu$m.

The first grating (V1) directly adapted processes used for X-ray grating fabrication and relied on a positive-tone e-beam resist, ZEP520A~\cite{zep}. As with other positive e-beam resists, the ZEP520A that was exposed during e-beam lithography processing was removed in the resist development step, leaving behind the unexposed regions to define the grating groove trace. The desired Si orientation required an off-axis cut to the Si that produced a narrow groove "pillar" on top of a broader base, as seen in Fig.~\ref{etchex}(a). The depth of the "pillars" etches more quickly than the width, causing the groove aspect ratio (the ratio of the groove width to the groove depth) to decrease to the point of mechanical instability when subject to extended KOH exposure. To limit the effects on groove structure from extended KOH etching, a higher ratio of exposed:unexposed area was pursued during e-beam exposure. When the ratio of exposed:unexposed grating regions grows too large, e-beam fogging can affect the resist, modifying the ultimate size of the groove features after development and degrading the fidelity of the residual resist~\cite{Chang15}. To both minimize the over-etching effects from the KOH exposure and avoid resist fogging, an exposed:unexposed ratio of $\sim$6:1 was used along with a controlled KOH etch, resulting in $\sim$1.5 $\mu$m Si plateaus at the apex of each groove in the 11.2 $\mu$m groove period. 

A second version (V2) of the CHESS grating was produced with smaller Si plateaus to provide a performance comparison to the V1 grating. The V2 grating used a negative e-beam resist, hydrogen silsesquioxane (HSQ)~\cite{hsq}, to circumvent the fogging issues encountered with the positive resist; the negative e-beam resist allowed the ratio of exposed:unexposed regions to be inverted during the e-beam write. Using HSQ, the V2 grating achieved a $\sim$20:1 ratio of groove period to feature width after e-beam exposure and resist development. The groove pattern was then transferred through the nitride layer and exposed to the KOH bath in the same manner outlined for the V1 grating. 

When the unexposed HSQ was removed after e-beam exposure, the remaining exposed HSQ formed an SiO$_2$-like material. This SiO$_2$-like residual resist etched more quickly than pure SiO$_2$ when exposed to KOH. Early test samples for the V2 grating demonstrated that the residual HSQ broke down quickly during the KOH exposure and potentially re-deposited on the sample during the etch process, resulting in a thin, anisotropic layer that adhered to the Si wafer surface. This layer did not significantly affect diffraction performance (see Section~\ref{chess_eff}), but was visible as a ``scum’’ layer on the grating surface that persisted through oxide removal and metal coating. 

The HSQ development also removed the unexposed resist across the entire wafer surface, including the wafer regions outside of the patterned area. The unmasked Si$_3$N$_4$ outside the patterned area was subsequently removed during dry-etch processing, leaving an unmasked Si surface during KOH etching. To minimize the excess etching of the unmasked Si outside the grating area and the physical area available for re-deposition of the developed HSQ during KOH etching, the V2 grating was diced down to a 70 mm $\times$ 60 mm rectangle prior to KOH exposure, leaving only the active grating area and a small unruled region to be used as a reflectivity sample (see Fig.~\ref{echelles}(b)). This piece was submounted onto a new 100 mm diameter by 1 mm thick wafer afterwards, which facilitates handling and mounting during testing. Both gratings received a gold coating deposited on top of a chromium adhesion layer. These layers were 30 nm Cr with 100 nm Au for V2 and 10 nm Cr with 50 nm Au for V2. This reduction in coating thickness minimizes the impact the coating has on grating performance after additional study indicated that a 50 nm thick coating was sufficient for achieving a reflective layer with little to no pitting.

\section{Efficiency Characterization}  \label{CHESS}
\subsection{Measurement Procedure} \label{chess_meas}

\begin{sidewaysfigure}
\centering
\captionsetup{width=\linewidth}
\includegraphics[width=\textwidth]{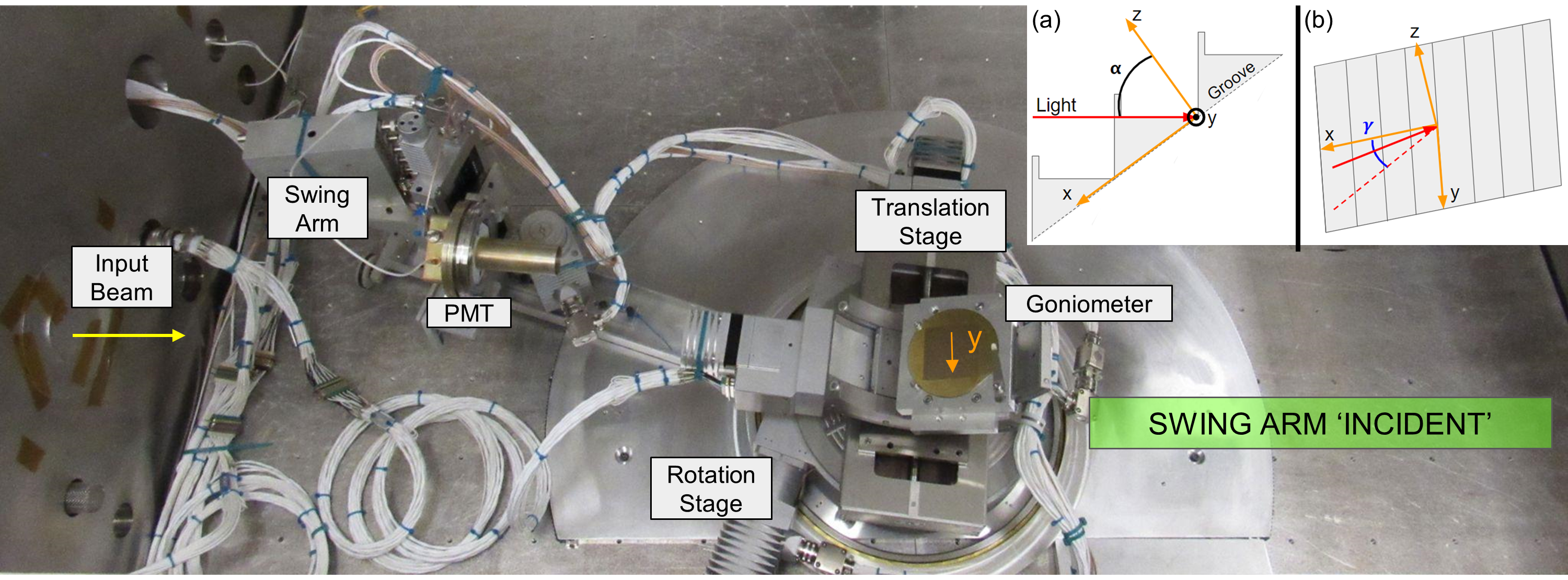}
\caption{The inside of the CU Boulder Square Tank with the V1 echelle installed. The PMT detector is shown on the swing arm and the swing arm and translation stage are located in the `reflected' position, where the incoming beam diffracts off of the grating. To sample the incoming beam, the swing arm is moved to the `incident' position, as indicated by the green box in the image, and the translation stage is used to move the grating and goniometer out of the path of the beam. The inset image in the top right shows the definitions of the grating $\alpha$ (Inset (a)), $\gamma$ (Inset (b)), and the grating coordinate frame. In the chamber, $\alpha$ is set by the goniometer, tilting the grating about the axis parallel to the grooves. $\gamma$ is set by the rotation stage, which rotates the grating azimuthally, disentangling the incoming and reflected beams.}
  \label{sqrtank}
\end{sidewaysfigure}

Efficiency measurements are performed in the University of Colorado Square Tank Chamber~\cite{France16,Moore16}, the layout of which is shown in Fig.~\ref{sqrtank}. The echelle gratings are mounted on a goniometer. This goniometer is attached to a horizontal translation stage that slews the grating in and out of the incoming beam, as well as a rotation stage. Two types of detectors are used to cover the entire bandpass - microchannel plate (MCP) detectors for $\lambda$ = 90 -- 120 nm and a photomultiplier tube (PMT) for $\lambda$ = 115 -- 160 nm. The PMT is a Hamamatsu R6836. Two different MCP detectors are used for different parts of the study: A Quantar 3391A and a custom MCP from Sensor Sciences LLC. The Quantar MCP is used for the peak order efficiency measurements. It reached the end 

\noindent of its life after that test was completed. The Sensor Sciences detector was purchased as a replacement but is still undergoing commissioning. In particular, it experiences high levels of ion backgrounds that arise when the swing arm stage is moved between the incident and reflected positions. This ion contamination has a significant decay time after motion and exists at levels that limit its accuracy for reflectivity measurements ($\sim$10\% of the reflected signal). The design and implementation of an ion repeller grid is in progress. This detector is used for follow up efficiency measurements and diffraction arc characterization. The light sources used for the measurements are an H-Ar fed hollow cathode discharge lamp for $\lambda$ = 90 -- 120 nm  and a Hamamatsu X2D2 deuterium lamp with an MgF$_{2}$ window for $\lambda$ = 115 -- 160 nm. These lamps feed an Acton VM-502 monochromator that is scanned across the bandpass to sample the grating efficiency at wavelengths of interest.

To facilitate the discussion on grating alignment and efficiency measurements, we have included an inset figure and additional pointers in Fig.~\ref{sqrtank}. The detector is mounted on the ``swing arm'' stage that scans azimuthally about the chamber center between the `incident' and `reflected' beams. The incident position allows the detector to directly measure the incoming beam without the grating in the optical path. This position is indicated by the green box in Fig.~\ref{sqrtank}. In the reflected position, the translation stage is used to move the grating into the optical path. The detector measures the diffracted beam by moving the swing arm around the chamber to this reflected position. This configuration is the one shown in Fig.~\ref{sqrtank}.

Several alignment steps are taken prior to starting efficiency measurements to ensure that the grating is in the correct geometric configuration. The inset of Fig.~\ref{sqrtank} provides context for the definition of the angles optimized during this alignment procedure. The inset (a) shows a side-on view of the grating, with the local axes defined having the y-axis parallel to the grooves and the z-axis aligned with grating normal. The y-axis is also labeled on the grating itself in the square tank image to assist in visualizing the alignment. We define $\alpha$ as the inclination angle between the incoming beam and grating normal, as shown in the inset figure. $\alpha$ is controlled in the chamber using the goniometer. For an echelle in Littrow, $\alpha$ is equal to the blaze angle of the facets (62.2$^{\circ}$ for the Si gratings). 

Inset (b) shows a top-down view of the grating, in a similar orientation to the square tank image, with the same axes defined. In this image, the dashed red line shows the projection of the incoming beam onto the grating surface. The angle between this projection and the x-axis defines $\gamma$. In Littrow, $\gamma$ = 0$^{\circ}$ results in the diffracted beam being directed back towards the incoming beam. Therefore, a $\gamma$ is introduced to disentangle the two, allowing the diffracted beam to be measured without blocking the incoming light. $\gamma$ is controlled in the chamber by moving the rotation stage. 

A grating is first visually aligned in the chamber using visible light fed through the monochromator, with the rotation stage set to $\gamma$ = 0$^{\circ}$. The grating is adjusted within its mount until the diffracted beam runs vertically and is coincident with the incoming beam aperture, putting the grating into Littrow and confirming that the grating facets are properly aligned relative to the mount. Once aligned, the swing arm is used to position the detector such that it is as close to the incoming beam as possible without vignetting (placing it in the reflected position), and then the rotation stage is moved to the point where the diffracted beam is centered on the detector setting $\gamma \approx$ 6$^{\circ}$.

With $\gamma$ defined, we proceed to vacuum alignments to refine $\alpha$. The Si substrate used for the echelles has an angular uncertainty of $\pm 1^{\circ}$. While we can set the grating to our desired $\alpha$, it may not be correct due to this uncertainty combined with uncertainties in our mount and alignment and that difference can impact grating performance. To ensure that we are illuminating the facets at the ideal angle for our setup, we first maximize the peak efficiency over a single order. The reflectivity across a 1.5 nm range is measured at several different $\alpha$'s and the angle with the highest peak reflectivity is selected for the remainder of
\begin{wrapfigure}{r}{0.5\textwidth}
\includegraphics[width=0.48\textwidth]{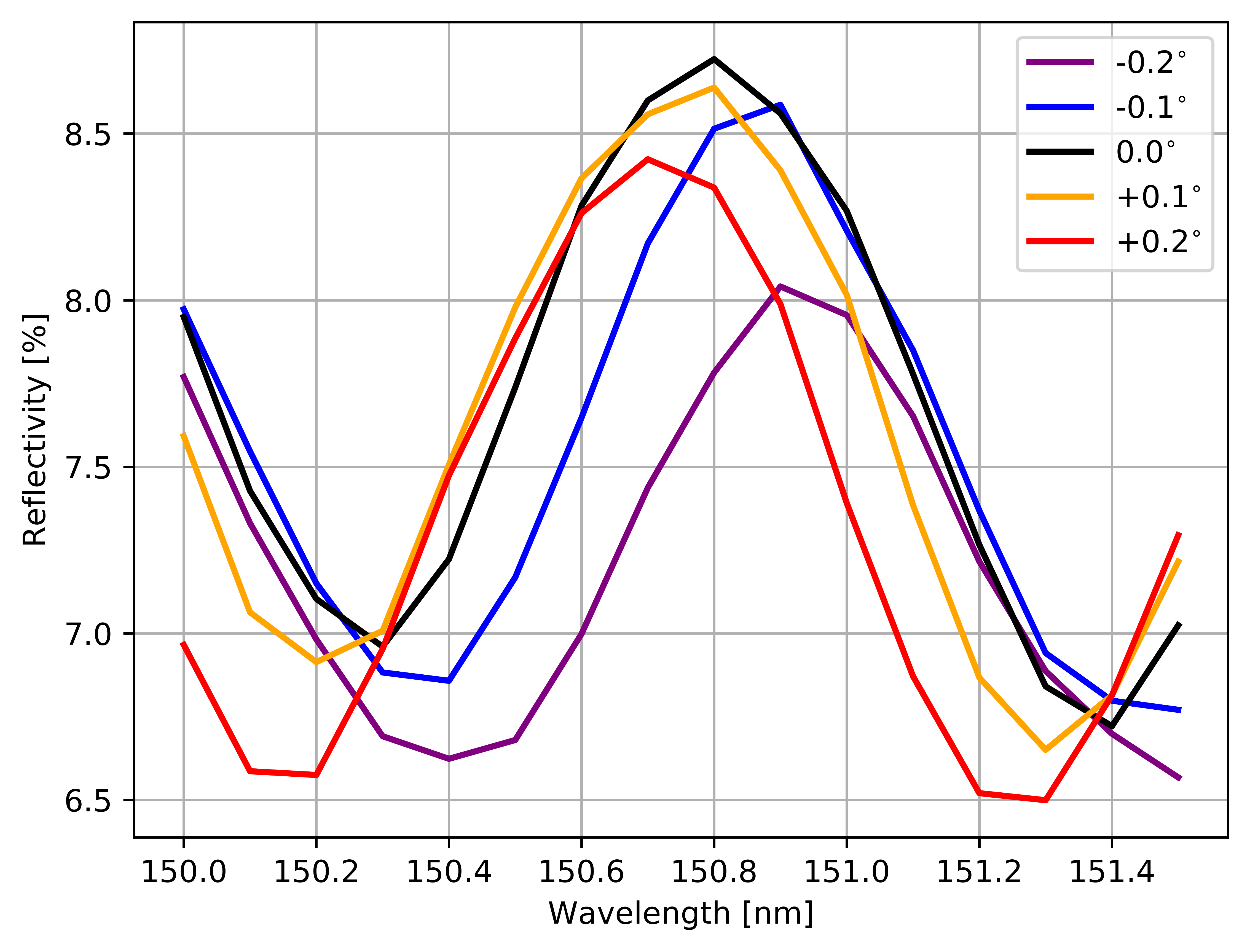}
    \caption{A demonstration of peak $\alpha$. Each curve shows a single order with a different $\alpha$, measured as an offset from the center peak value by the amount listed in the legend.}
  \label{gonipk}
\end{wrapfigure}
the efficiency study (Fig.~\ref{gonipk}). We found that deviations from the expected $\alpha$ were at most 0.2$^{\circ}$ relative to the nominal $\alpha$ = 62.2$^{\circ}$. A perfectly aligned grating would have a diffraction arc running vertically. As a check to our alignment, we can observe several echelle orders using the MCP detector and measure a tilt based on their positions. We find values on the order of 0.2 -- 0.3$^{\circ}$, we do not expect offsets of that magnitude to impact efficiency significantly.

To check that we are sampling the peak order efficiency over the entire bandpass, we repeat the 1.5 nm wide scans at three different wavelength regions. We assume the maxima of these peaks correspond to the blaze wavelength for integer order solutions to the grating equation and use their values to produce an equation consistent with the observed distribution. Performing 1.5 nm wide scans across the entire bandpass is a time consuming process. Instead, we use this analytic grating equation fit to generate a list of the remaining peak wavelengths, which we then sample when measuring the peak order grating efficiency.

\indent To obtain an efficiency, we scan across 16 wavelengths in total (four for the MCP, 13 for the PMT, with one wavelength overlapping) spanning 97 - 160 nm. A dark wavelength is included at the beginning and end of these scans. This wavelength is chosen to be a value where the monochromator efficiency drops below the backgrounds, serving as a proxy for the monochromator scatter and detector noise. We measure count rates at the incident and reflected positions, the ratio of these values provides the grating efficiency. The grating is then removed from the setup and reinstalled in a configuration where the reflectivity of the gold coating can be measured at the same $\gamma$. For the Si gratings, the coating reflectivity is measured from an unruled portion of the grating substrate. The CHESS echelle received a 50 nm thick gold coating, which was applied by the vendor. We do not have the benefit of an unruled portion of the grating to use to test reflectivity and so we instead infer this value by measuring the reflectivity of an unruled substrate that is otherwise identical to the grating and that received the same coating prescription. Differences could exist between these coatings but based on results presented in this work (Section~\ref{inst_perf}) we do not expect that this possible discrepancy is driving the efficiencies measured in this work.

\subsection{Efficiency Results} \label{chess_eff}
\begin{figure}[t]
\centering
\includegraphics[width=\textwidth]{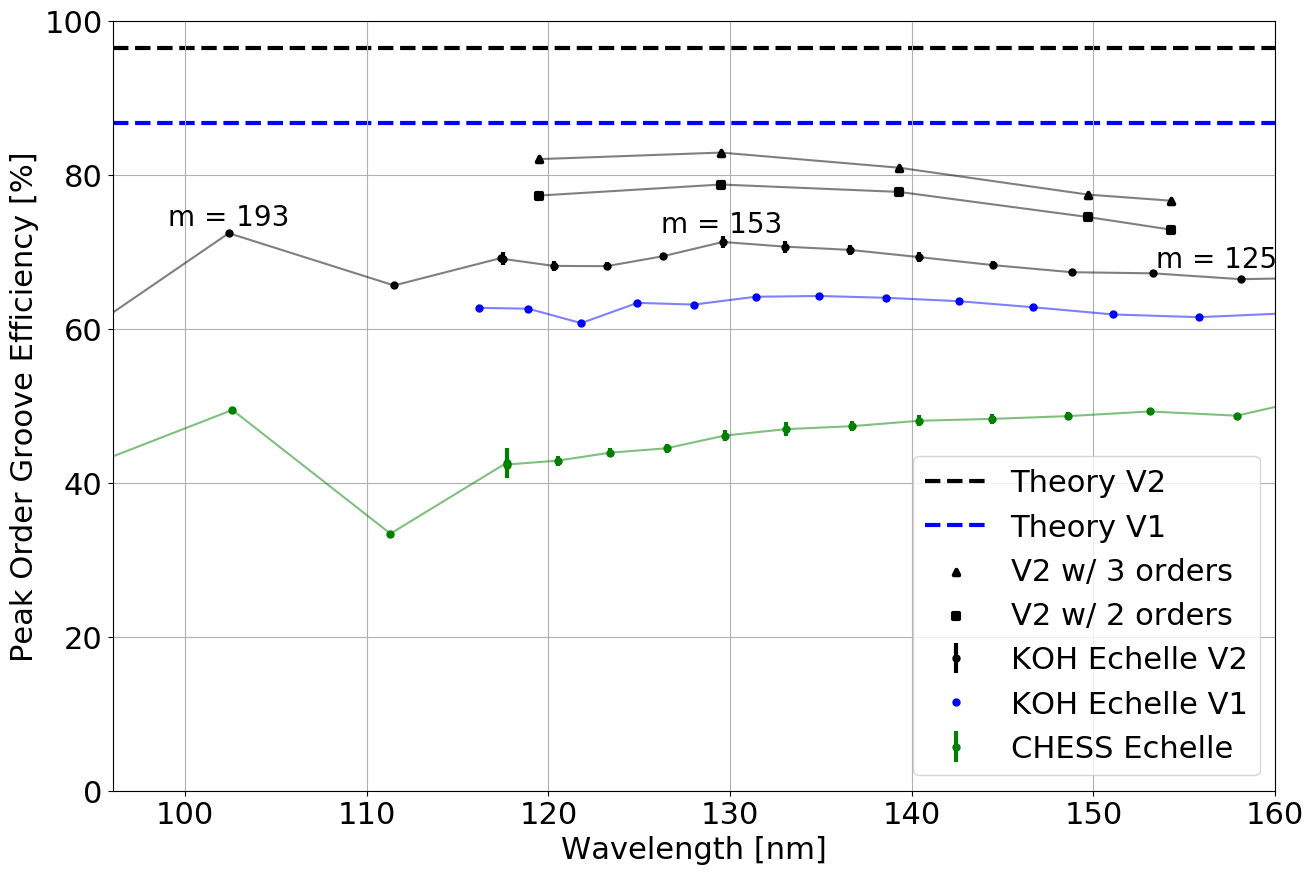}
    \caption{The groove efficiency of the 0.4 $\mu$m-wide plateau echelle (V2; black), the 1.5 $\mu$m-wide plateau echelle (V1; blue), and the mechanically ruled CHESS echelle (green). Peak order groove efficiencies are not continuous functions, they are discreet points at the apex of each individual order. Lines have been included for each set of points to aid in clarity only, not to imply a continuous distribution in wavelength. Off peak wavelengths reach similar, or larger, groove efficiencies only after including contributions from two orders. The corresponding orders for a selection of points for the V2 echelle have also been included. The dashed lines show the theoretical efficiencies for the two lithographic gratings assuming only geometric losses. We demonstrate how the groove efficiency of the V2 grating improves as the second (square points) and third (triangle points) brightest orders are included in the efficiency measurement.}
  \label{groove_eff}
\end{figure}

The resulting peak order groove efficiencies of both lithographic gratings as well as the mechanically ruled CHESS grating are shown in Fig.~\ref{groove_eff}. When comparing the groove efficiencies between the gratings, we find a significant improvement in the performance of the Si gratings. In particular, the V2 grating demonstrates a $\sim$50\% increase in efficiency (20\% absolute change) across the bandpass. The short wavelength measurements were taken using the Quantar MCP prior to it being decommissioned. The declining performance of this detector lead to severe gain sag during the measurements. This effect is evident in the systematic variability of measurements below 120 nm. These values do vary about a mean that is consistent with the long wavelength results.

The square and triangle points in Fig.~\ref{groove_eff} show the theoretical maximum efficiency of the Si gratings assuming only geometric losses to their plateaus. These values are equal to 86.7\% for the V1 grating and 96.5\% for the V2 grating. Ideally, all of the remaining light would be directed to a single order, achieving the maximum possible groove efficiency. As discussed in Section~\ref{intro}, natural imperfections and nonidealities in all grating fabrication methods and coating processes result in gratings that deviate from this idealized performance, directing light away from the blaze order into neighboring orders and interorder scatter. It is helpful to characterize how this light is distributed for several reasons including providing a baseline metric as this technology continues to develop and to inform instrument design, since light diffracted into neighboring orders can still be imaged on the detector if designed properly.

We provide a brief demonstration of this using the V2 grating installed in the same configuration. We measure the two brightest neighboring orders about the peak, and add them onto the peak order groove efficiency. Additional orders would have been included but the geometry of the square tank limited the range the detector could cover. These measurements took place after we had installed the new MCP detector (Section~\ref{chess_meas}) and the grating had been removed from the chamber. Therefore, we treat them as a demonstration of the evolution of total groove efficiency. This MCP detector has a CsI photocathode, improving its red sensitivity. We ran this test at longer wavelengths because the D$_{2}$ lamp has a higher signal, which helps reduce the impact of the ion backgrounds.

The result of this measurement is shown in Fig.~\ref{groove_eff}, demonstrating the change in the groove efficiency as the additional orders are included in the diffracted counts. While a vast majority of the light does reside in the peak order, the addition of these neighboring orders results in a non-negligible increase in the overall efficiency. With proper instrument design and detector selection, groove efficiencies on the order of 80\% could be achievable with these new echelles in their current state. The remaining $\sim$20\% between the three order line and the theoretical maximum is expected to reside in additional outer orders, given the predicted low scatter performance of these gratings. We can explore this in more detail by studying the complete diffraction arc.

\section{The Diffraction Pattern} \label{diffpatt}
\subsection{Diffraction Arc Scans}\label{arcs}
\begin{figure}[b]
\centering
\includegraphics[width=0.99\textwidth]{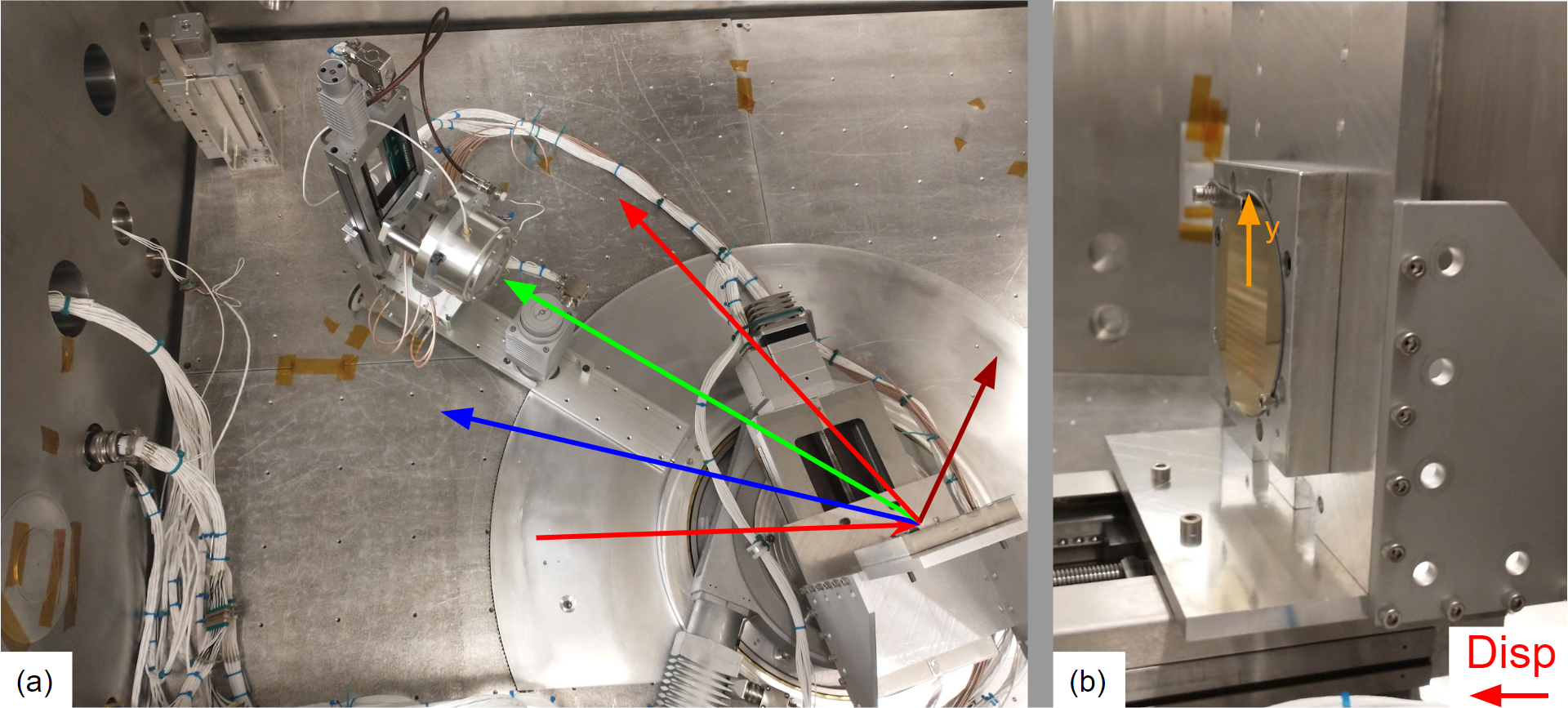}
\caption{(a) A top down view of the square tank in the configuration used for the diffraction arc scans. In this orientation, the dispersion axis runs along the swing arm's path so that the detector can scan along the entire diffraction arc. (b) An image of the V1 grating mounted in this configuration. The dispersion axis runs horizontally in this image. This is  a 90$^{\circ}$ rotation about grating normal relative to the orientation in Fig.~\ref{sqrtank}. We have labeled the grating's y-axis as a reference to the grating coordinate system in that figure.}
\label{config2}
\end{figure}
To quantify the KOH-etched echelle diffraction arc, we remove the goniometer from the square tank chamber and mount the grating to a vertical plate in a new configuration (Fig.~\ref{config2}). In this orientation, the grating is rotated 90$^{\circ}$ about grating normal from the original setup, such that $\alpha$ now runs azimuthally about the chamber, meaning the diffraction arc runs along the same axis as the swing arm stage and the swing arm can be used to scan the full diffraction pattern. With the grating hard-mounted to the plate, the grating is fixed at $\gamma$ = 0. We are unable to operate the grating in Littrow in this configuration because the detector would block the incoming beam when trying to measure the blaze order. To avoid this, we increase $\alpha$ by $\sim$9\adeg, which was the smallest angle where we can scan all of the observable orders without vignetting the beam. While this change impacts the diffraction pattern, the changes are slight and our goal with this process is to quantify where the light is diffracting in general, not compare efficiency levels in specific orders.

\begin{figure}
\centering
\includegraphics[width=\textwidth]{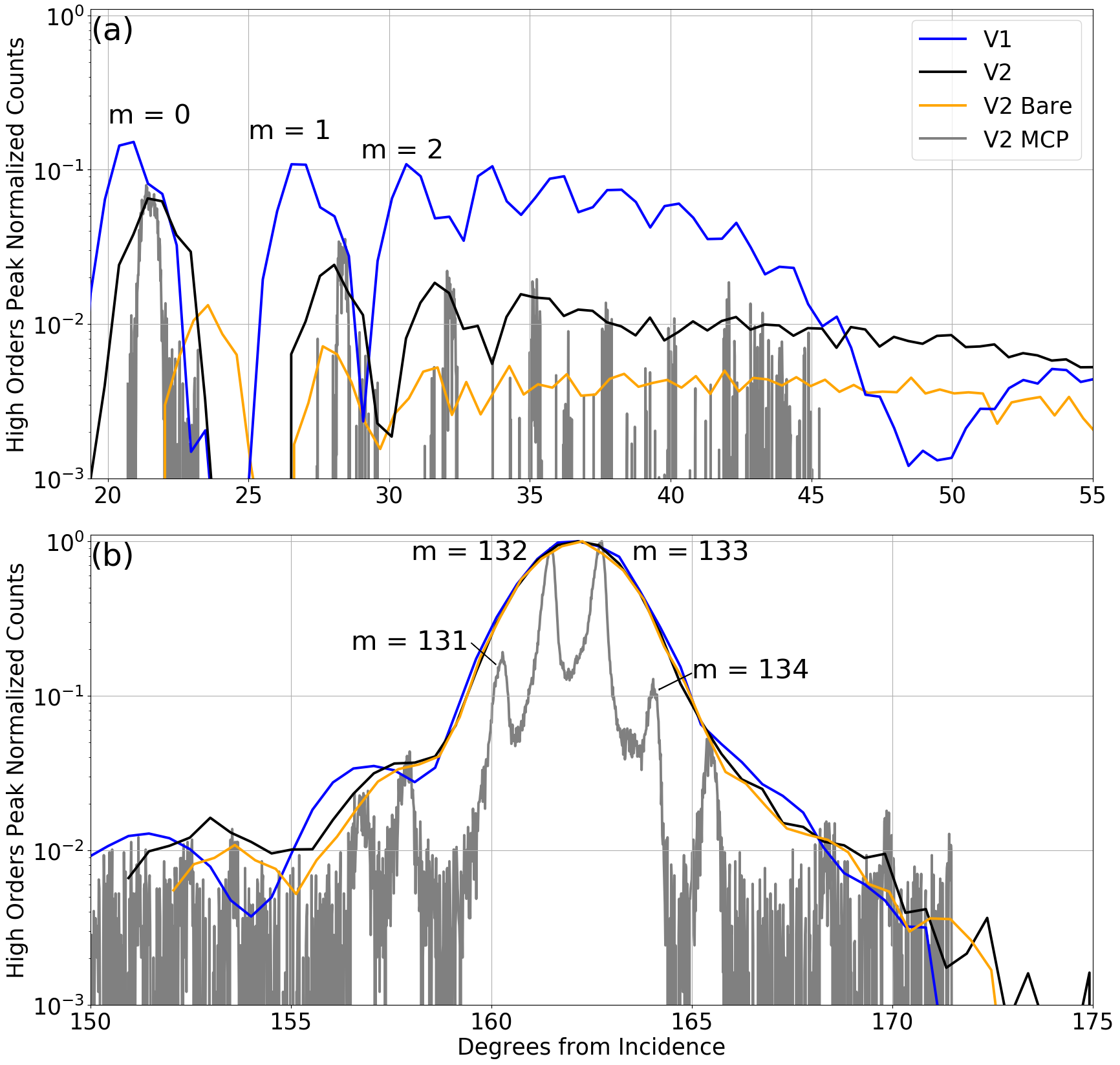}
    \caption{The diffraction arcs of the Si gratings for the low (a) and high (b) order diffraction arcs. The curves in both plots are normalized to the peak high order value (near 162$^{\circ}$). The blue line shows the observed pattern for the gold coated V1 grating. The V2 grating is shown both before (orange) and after (black) the gold coating was applied. The gray line shows the scan of the gold coated V2 grating taken with the new MCP. The x-axis tracks the swing arm angle relative to the incident position. In that coordinate system, the blaze order would be at 180\adeg if the grating was in Littrow. The observed peak orders are seen at $\sim$162\adeg due to our rotation out of Littrow to cover the entire diffraction arc.  A selection of orders have been labeled in both plots for reference.}
  \label{sa_scans}
\end{figure}

Four diffraction arc scans are shown in Fig.~\ref{sa_scans}, normalized to the peak counts. All scans are at 150 nm. The scans were primarily taken using the PMT detector, with imaging of the individual diffraction peaks taken using the Sensor Sciences MCP. For each grating, we find both a high order diffraction pattern just out of Littrow (150\adeg --170\adeg from incident) and a second diffraction pattern closer to the incident position (20\adeg --60\adeg). This second pattern arises from diffraction off of the flat Si plateaus, which have the same groove density and therefore generate a secondary grating that is effectively optimized for m = 0 (specular reflection) with a non-zero efficiency in nearby orders. Two PMT scans of the V2 grating are included. These were taken before and after the gold coating was applied. We attempted to measure the groove efficiency of the V2 grating pre-coating but did not have a usable sample of unruled substrate to reliably measure a reflectivity. The V1 grating has the largest plateaus. This results in it having the largest fraction of total power diffracted into low orders relative to the peak high order counts. The application of the gold to the V2 gratings widens the plateaus slightly, resulting in more power being directed to low orders. While this trend is visible in Fig.~\ref{sa_scans}(a), as the amount of power in the low orders increases across the three measurements, the comparison is not perfectly 1:1 due to differences in substrate reflectivities.

For all three PMT scans of the high orders, we see a primary peak near 162\adeg and additional ringing at lower angles (158\adeg and 153\adeg). Similar to the low orders, this ringing is most pronounced in the V1 grating and the least pronounced in the bare V2 grating. This trend presumably results from coating differences, as the application of a coating will round out the otherwise sharp Si facets, resulting in more power diffracted into off-blaze orders. As expected from the efficiency measurements, the MCP scan of the coated V2 grating shows four individual echelle orders that all have appreciable efficiencies. The peak order is evenly split between two of those orders, which is expected given that the wavelength for this scan (150 nm) has a non-integer order solution to the grating equation in this geometry. 

The ringing seen in the high order curves occurs in the direction favored by the out of Littrow orientation, it is likely that the ringing asymmetry is present due to this rotation and that, in Littrow, the profile would look symmetric. The combined efficiency of the two peak orders in the MCP scan is $\sim$70\%, in good agreement with Fig.~\ref{groove_eff}. This indicates that the grating orientation is not significantly impacting the efficiency of the peak order(s). Therefore, while the asymmetry we see in the diffraction pattern is likely due to the grating orientation, we do not expect that this power would be redirected to the blaze order in Littrow. Instead, we would expect to see this power shared between symmetric orders at the same angular spacing on either side of the blaze order.

We confirm the geometric predictions shown in Fig.~\ref{groove_eff} by calculating a total groove efficiency from the PMT scans for these low and high orders. We measure a total grating efficiency using the total counts in each region as well as the reflectivity of the gold coating at the corresponding $\alpha$'s for the two regions. Table~\ref{grvreg} lists the measured groove efficiencies for the low and high orders of the gold coated V1 and V2 gratings. These values include light present as interorder scatter. The table also lists the expected groove efficiency based off of the geometry and the total efficiency, which serves as a check that we are accounting for all of the light. The error bars on these total efficiency measurements come from standard error propagation methods for combining the errors present in the diffraction arc scans and gold reflectivity measurements. We find total efficiencies consistent with 100\% to within the error bars for both gratings. This demonstrates that, while we may not have a sub-percent accuracy in our light accounting procedure, we can rule out that our observed discrepancy between the measured and geometrically-predicted efficiencies is not due to missing light in the system, in line with the expectation that the discrepancy lies in scatter and the off-peak orders. 

For both gratings, we find that the measured low order groove efficiency is larger than the geometric expectation. For the V1 grating, this discrepancy can be explained by a widening of the plateaus due to the application of the gold coating. Assuming the worst case, where 0.13 $\mu$m of material is added to either side of the plateau, we would expect the V1 plateaus to reflect 15.7\% of the light after coating. This is compared to the measured 14.5\%, showing that this increase is consistent with some amount of widening due to the coating. For the V2 grating, this same issue would exist but we observe a larger shift in power towards these low orders than would be expected if an additional 0.12 $\mu$m is added to the total plateau width. In this case, the additional width would only lead to 4.6\% of light reflecting off the plateaus, below the observed value of 5.7\%. The cause for this increase is not currently known and will require a more detailed study of how the application of a coating impacts the shape and size of the facets.

\begin{table}
\centering
\caption{Total Groove Efficiencies [\%]}
\label{grvreg}
\begin{tabular}{| c | c | c | c | c |}
\hline
\multirow{2}{*}{Grating} & \multirow{2}{*}{Low Orders} & Low Orders &  \multirow{2}{*}{High Orders} &  \multirow{2}{*}{Total} \\[-4pt]
& & (Geometric) & & \\ \hline
V1 & 14.5 $\pm$ 0.2 & 13.3 & 85.1 $\pm$ 1.5 & 99.6 $\pm$ 1.5 \\ \hline
V2 & 5.7 $\pm$ 0.1 & 3.6 & 95.1 $\pm$ 1.6  & 100.7 $\pm$ 1.6 \\ \hline
\end{tabular}
\end{table}

Despite the observed redistribution of light to low and off-blaze orders, these KOH-etched gratings still outperform traditional mechanically ruled gratings. Continued development of this technology could further improve this performance and concentrate the outlying energy more centrally about the blaze order. Table~\ref{grvreg} shows that the observed diffraction arc is consistent with the geometrically predicted distribution to within 1-2\% and a majority of that light is concentrated within one or two peak orders. While it looks like a most of the light outside of the peak order is concentrated in outer orders, it's difficult to separate scatter from square tank noise and spot structure. We can better understand scatter and demonstrate the benefits of this new technology through in-instrument comparisons.

\subsection{Spot Structure} \label{sptstr}
\begin{figure}[b]
\centering
\includegraphics[width=0.75\textwidth]{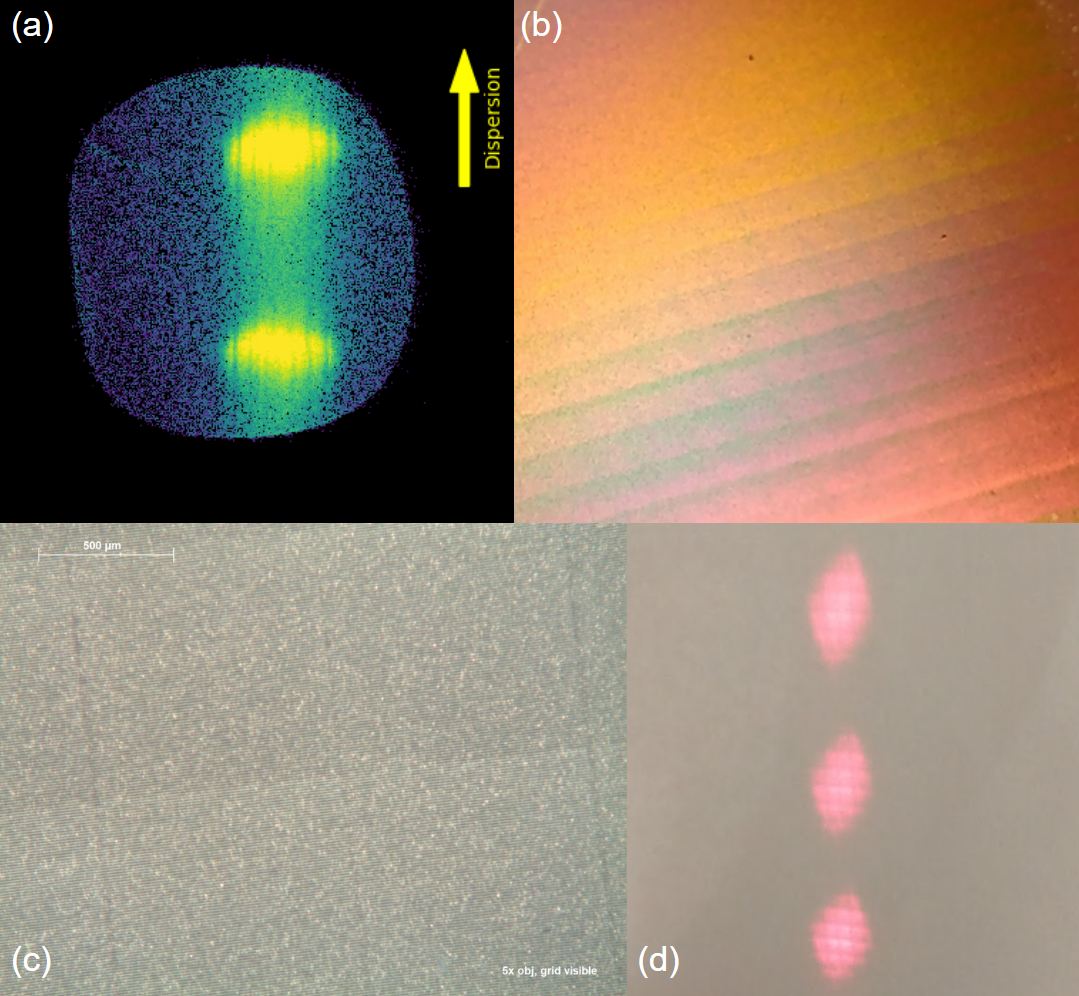}
    \caption{(a) The streaked pattern observed in the square tank spot, in this case at 150 nm. The lines run along the dispersion direction and repeat across the cross dispersion direction. (b) The write area grid effect observed along the surface of the V2 grating. A similar pattern is seen on the V1 grating. (c) Four of these write area cells observed under a microscope, where the narrow write area edges are observable as faint black lines. (d) The observed grid pattern using the laser beam in the bench-top setup. In this case a full grid pattern is seen.}
  \label{spotgrid}
\end{figure}

Preliminary observations of the diffracted spot off of the Si gratings using the MCP detector showed the presence of streaks running along the dispersion axis and repeating along the cross dispersion axis (Fig.~\ref{spotgrid}(a)). The size of these features is consistent with the e-beam tool write field size and visual inspection of the surface at multiple angles reveals a grid pattern that matches this effect (Fig.~\ref{spotgrid}(b)). Observing these boundaries under a microscope shows that the grid manifests as narrow dark features (Fig.~\ref{spotgrid}(c)). These features likely arise due to small discontinuities between neighboring write areas.

The microscope image shows features that subtend a smaller area than might be expected based on the size of the features observed in the square tank exposure. We use a bench-top setup to study this in more detail, illuminating the grating with a red laser ($\lambda$ = 635 nm) with a spot large enough to subtend a roughly 4 $\times$ 4 grid of the square regions on the grating surface. Near the grating, the spot shows lines similar to what was observed in the square tank (Fig.~\ref{spotgrid}(d)). In this case, the pattern is seen in both the dispersion and cross dispersion directions. We likely did not see the horizontal gaps in the square tank image due to the nonzero $\Delta \lambda$ bandpass of the monochromator. Having a small range of wavelengths would smear the diffracted spot and hide these features.

\begin{figure}[b]
\centering
\includegraphics[width=\textwidth]{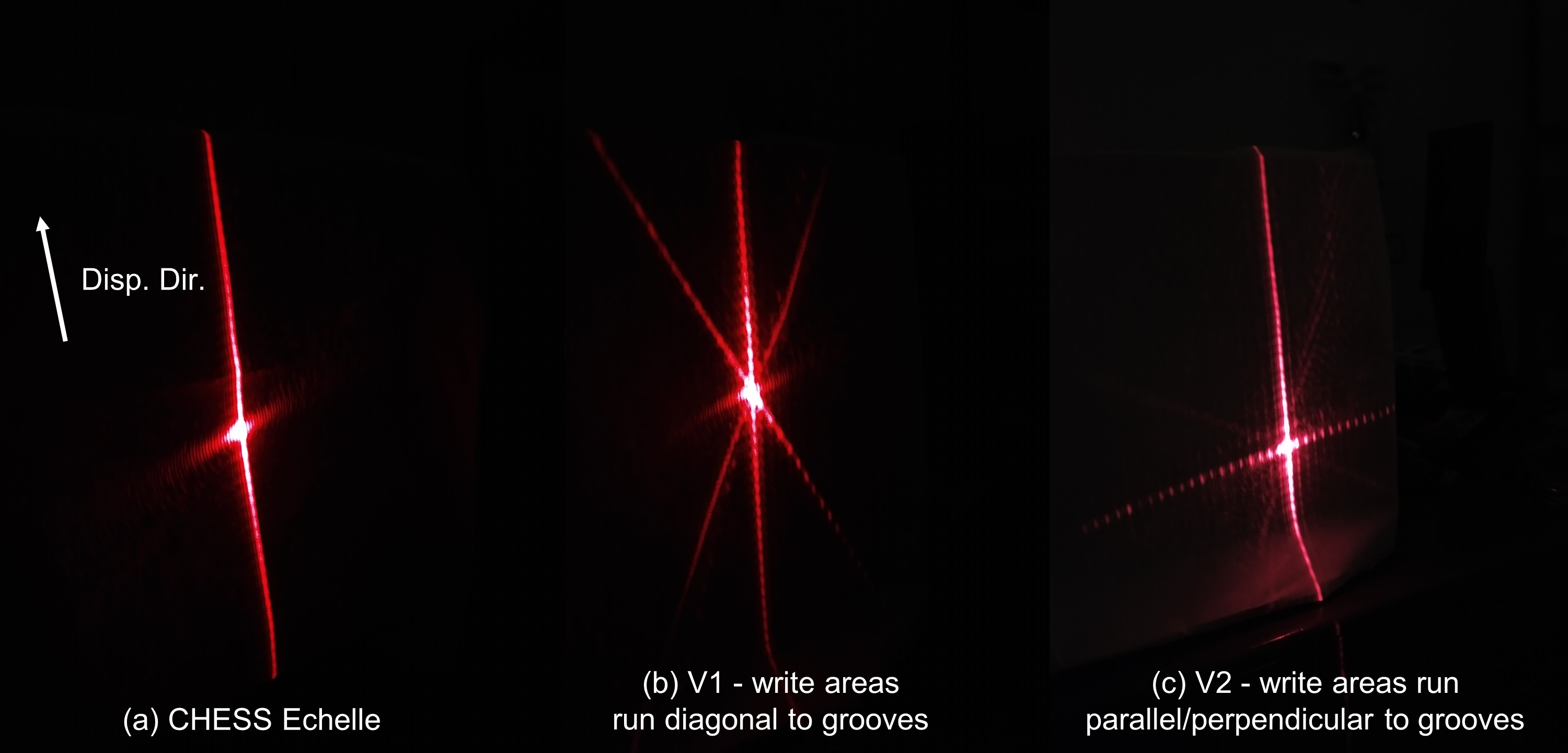}
    \caption{(a) The peak order diffracted spot from the CHESS echelle, showing the expected profile for an echelle grating. (b) The peak order diffracted spot from the V1 echelle. Here the e-beam write areas run diagonally relative to the grooves, so an X-shaped square aperture diffraction profile is observed superimposed on the grating spot. (c) The peak order diffracted spot from the V2 echelle. Here the e-beam write areas run parallel/perpendicular to the grooves, so a cross-shaped square aperture diffraction profile is observed superimposed on the grating spot.}
  \label{farfield}
\end{figure}

Further from the grating surface, the Si gratings show the typical echelle diffraction profile, but each individual order additionally contains a secondary diffraction pattern. This is shown in Fig.~\ref{farfield} for the three gratings studied in this work. The mechanically ruled grating acts as a control for the V1 and V2 gratings. The write area cells on the V1 grating run diagonally relative to the grooves, this results in an X-shaped pattern relative to a diffracted spot that otherwise looks similar to the mechanically ruled grating. The V2 grating has write area cells that run parallel/perpendicular to the grooves resulting in a secondary diffraction pattern that runs perpendicular to and along the grating dispersion axis.

This secondary pattern matches that of a square aperture, indicating that these individual write regions act as independent apertures. At FUV wavelengths, the grating-to-detector distance in the square tank falls in the near field diffraction regime. The near field diffraction pattern of a square aperture can be solved analytically. Following the notation of Goodman~\cite{Goodman05}, the intensity of a 1D square aperture of width $l$ at a distance $z$ from the aperture is:

\begin{multline}
I(z) = \frac{1}{\sqrt{2}} \left[C\left(\sqrt{2N_F} (1 - 2z/l)\right) + C\left(\sqrt{2N_F} (1 + 2z/l)\right) \right] \\ + \frac{j}{\sqrt{2}} \left[S\left(\sqrt{2N_F} (1 - 2z/l)\right) + S\left(\sqrt{2N_F} (1 + 2z/l)\right)\right]
\label{int}
\end{multline}

\begin{figure}[t]
\centering
\subfigure{\includegraphics[width=0.48\textwidth]{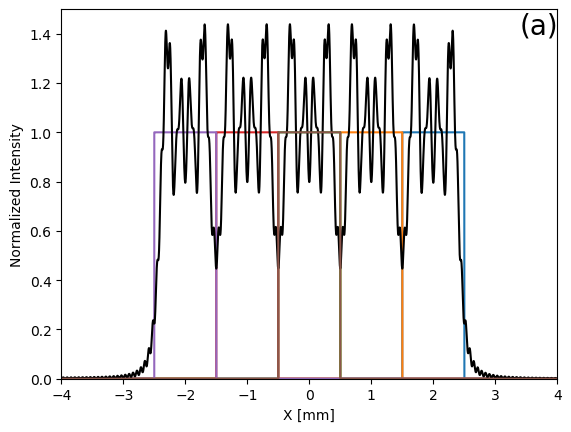}}
\subfigure{\includegraphics[width=0.49\textwidth]{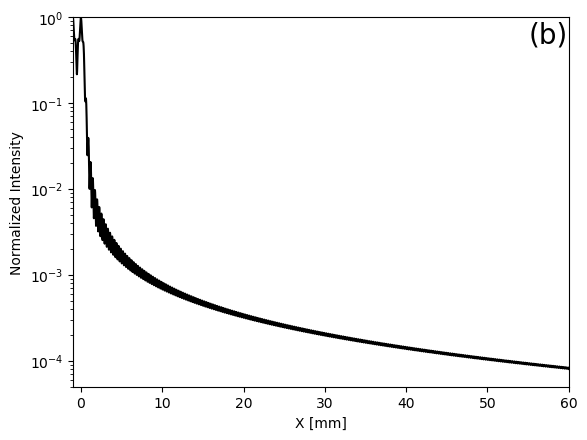}}
\caption{(a) The 1D near field diffraction pattern for five square apertures for N$_F$ = 4.0 (black line), which is the average value for the square tank in the FUV. The five colored rectangles show the initial square apertures, which have effectively no edge separation between their neighbors. This near field diffraction pattern forms periodic troughs like we observe in the square tank spot (Fig.~\ref{spotgrid}(a)). (b) The falloff of the total diffraction profile for an assembly of 100 square apertures, simulating the observed edge effects from the side of the echelle grating. The apertures run from x = -100 to 0 mm and only the peak at x = 0 is shown.}
\label{simspot}
\end{figure}

\noindent Where C and S are the Fresnel integrals and N$_F$ is the Fresnel number, which is the metric used to roughly delineate the near and far field regimes. When observing light with wavelength $\lambda$ , N$_F$ is defined as:
\begin{equation}
N_F = \frac{(l/2)^2}{\lambda z}
\end{equation}
N$_F$ $\gtrsim$ 1 is typically considered the near field. At square tank distances, N$_F$ = 3 -- 5 across the CHESS bandpass. A simulated idealized diffraction pattern for five square apertures is shown in Fig.~\ref{simspot}(a), calculated using Equation~\ref{int}. Note, this assumes a perfectly coherent spherical wavefront is incident on the apertures and is only meant to demonstrate that the effect can generate notable periodic gaps in the imaged spot, like we see in the square tank. The square tank image is generated from a less ideal light source so will not agree exactly. This analysis, along with Table~\ref{grvreg}, indicate that the write region boundaries do not cause a significant loss of light ($\lesssim$ 1\%). Instead, the observed gaps only appear large because of these diffraction effects. Their impact will be more important on instrument performance, where bright lines could display the effects shown in Fig.~\ref{farfield} if N$_F$ is sufficiently small.

The Fresnel number for these square apertures in the CHESS instrument range from 1.53 (at 100 nm) to 0.96 (at 160 nm). Again using Equation~\ref{int} to generate the combined intensity profile for 100 square cells, which is the number that would exist across a full echelle in the instrument, we estimate the impact this effect would have on the performance of CHESS (Fig.~\ref{simspot}(b)). For the idealized instrument, a feature that is 50 $\mu$m away from the edge of a spot on the detector will overlap with either a neighboring order in the cross dispersion direction or lie beyond the CHESS resolving power requirement (R $\sim$ 120,000) in the dispersion direction. 50 $\mu$m in image space is equivalent to 50 mm in echelle space. At this distance, the collective write area diffraction effects are expected to be at most 1.0 $\times$ 10$^{-4}$I$_{0}$, where I$_{0}$ is the peak line intensity. In both directions, this diffraction effect would only have a modest, if any, impact on the CHESS results, given that its targets lack bright emission lines. A study concerning particularly bright emission features may need to consider these effects. Nonetheless, the presence of this diffraction pattern warrants further study to better understand its impact. Additionally, multiple square apertures will collectively exhibit a grating like behavior. We would not expect this pattern to negatively impact performance given that the efficiency envelope of the grating would be defined by the aperture diffraction geometry of a single square aperture "groove"~\cite{Schroeder99}.

\section{In-instrument Performance} \label{persec}
\subsection{CHESS Demonstration} \label{inst_perf}
We can further demonstrate the gains provided by the KOH etching process by comparing how these new echelles perform in the CHESS instrument relative to the original mechanically ruled grating. The V1 grating was chosen over the V2 grating due to damage present on the surface of the V2 grating that we were concerned could impact the analysis. The CHESS payload was reassembled from storage, using the same flight cross disperser and flight MCP detector. The payload was aligned in the CU Boulder Long Tank facility, which is a large vacuum collimator. See previous CHESS references for complete details about Long Tank operation, payload alignment, and instrument operating principles~\cite{Hoadley14,Kruczek17,Kruczek18}. Both the mechanically ruled and V1 gratings are installed using the same mount. A base plate is used to ensure that the V1 grating sits at the same height within the mount as the CHESS echelle. The payload is initially illuminated with visible light to align the V1 grating by hand, prior to vacuum operations.

\begin{figure}[t]
\centering
\includegraphics[width=\textwidth]{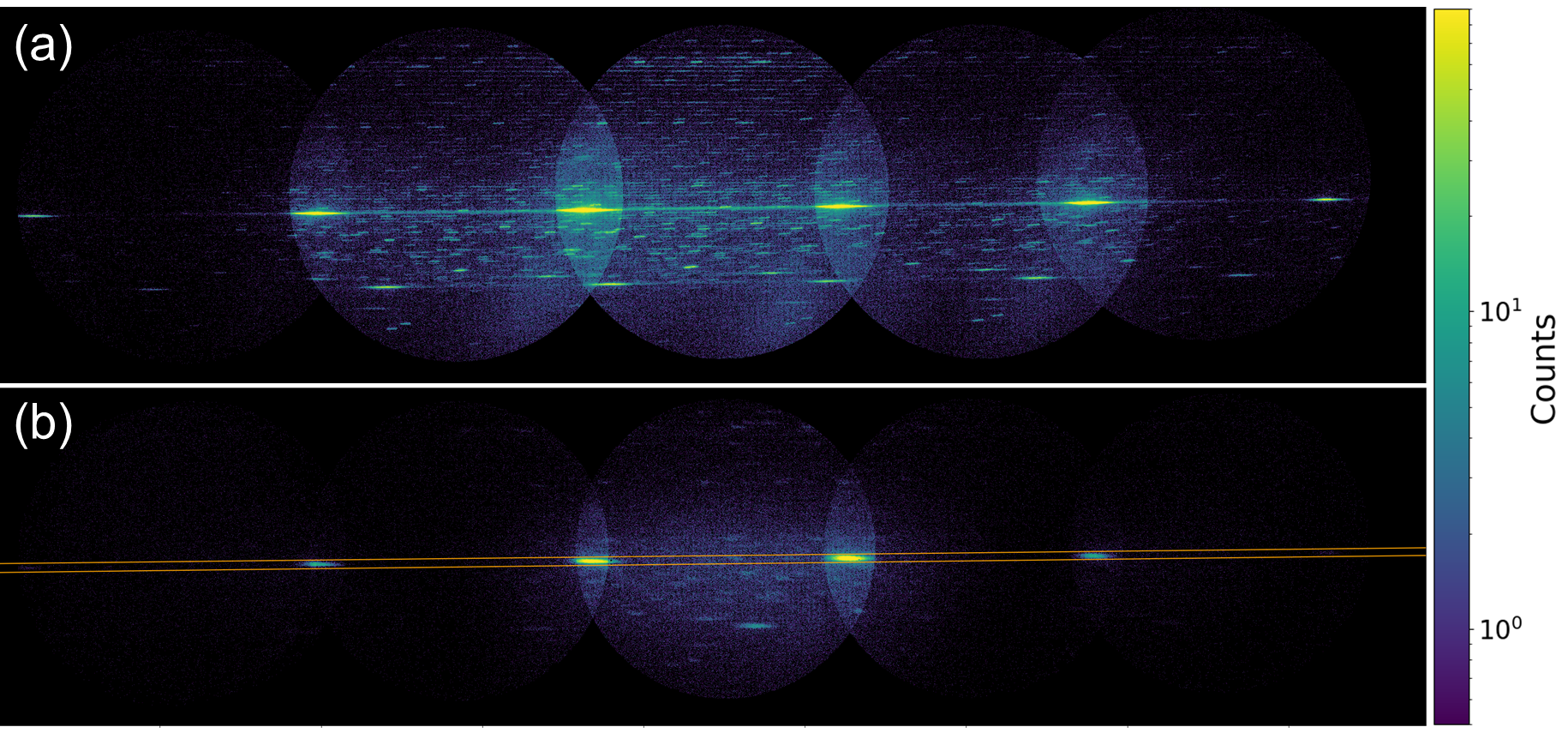}
\caption{Stitched echellograms of the mechanically ruled echelle (a) and V1 grating (b). Each circle in the image is an individual detector exposure, generated by tilting the cross disperser along the echelle dispersion axis to cover the range of echelle orders. The orange lines shown in the lower image contain the band of Ly$\alpha$ orders. The region between these lines was used to generate the 1D extracted spectrum shown in Fig.~\ref{1dinst}. The peak order efficiency gains from Fig.~\ref{groove_eff} and the improved scatter performance of the KOH-etched grating can be clearly seen here, where a majority of the light is concentrated in the central two orders. This is compared to the more numerous bright orders and prominent scatter feature seen for the mechanically ruled grating in (a). }
\label{ininst}
\end{figure}

The cross disperser is mounted on a set of actuators, which are primarily used for focusing the instrument but can also position the spectrum on the detector. We use this second capability to generate a map of the echelle diffraction arc for both gratings over the cross disperser's range of motion. The resulting images are shown in Fig.~\ref{ininst}. Each circle in that image is an individual detector exposure. The central circle in each image contains the peak orders, this is the `flight-like' alignment of the instrument. The cross disperser is scanned in both directions along the dispersion axis to capture the neighboring orders. Each exposure is stitched together to form the final image. The most prominent feature in each image is the string of Lyman $\alpha$ (Ly$\alpha$; 121.6 nm) orders, which are contained within the orange lines in the stitched V1 grating echellogram. 

\begin{wrapfigure}{r}{0.45\textwidth}
\includegraphics[width=0.45\textwidth]{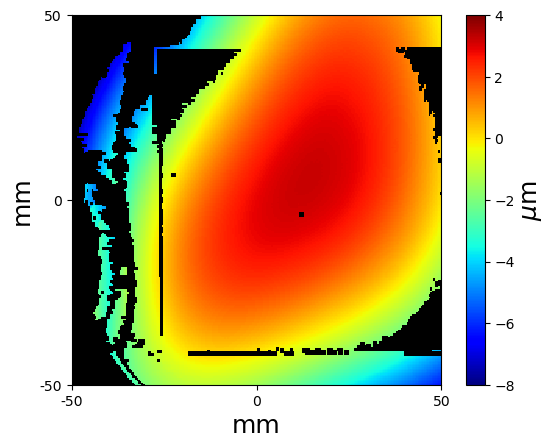}
    \caption{The measured curvature of the V1 grating, taken with a Fizeau interferometer using a 632.8 nm laser. This surface shape induces the defocus that we see for the grating in Fig.~\ref{ininst}(b). The dark regions show patches that were either too low signal or fall out of the measurement range of the interferometer. This image serves as a qualitative demonstration of the grating curvature and these regions do not impact that interpretation.}
  \label{curvy}
\end{wrapfigure}

We are unable to properly focus the V1 grating, which is evident in the larger order widths and the lack of obvious emission lines outside of the brightest features, since the signal from dimmer lines is spread over a larger area. Subsequent analysis using a Fizeau interferometer found that this is due to a curvature in the grating substrates (Fig.~\ref{curvy}). The V2 grating was also tested and demonstrated a similar curvature and so we do not plan on testing it in-instrument. Diagnosing and correcting this behavior is planned as a future point of study, with the likely solution simply requiring the use of thicker substrates. Despite this issue, we can see the benefits provided by the lithographic fabrication process. The first benefit is the previously discussed improved peak order efficiency of the Si echelle, which is evident in the concentration of the spectrum towards the center of the stitched echellogram. The second benefit is observed in the interorder scatter. Echelle interorder scatter follows a unique path across the detector, linking diagonally between two identical wavelengths in two neighboring orders~\cite{stisbad}. This form of scatter is noticeably more prominent in the mechanically ruled echellogram. The defocus of the V1 echellogram could complicate this interpretation.

To confirm these observations, we perform 1D extractions of the Ly$\alpha$ diffraction profiles for each grating. For both gratings, an extraction region is defined that runs along the echelle dispersion direction. This region is shown as the orange lines in Fig.~\ref{ininst}(b). A similar region is defined for the mechanically ruled echelle but we chose to omit it from the plot to provide better clarity of the scatter profile. We generate a 1D spectrum by collapsing each extracted region vertically. A coarse background correction is performed by defining identically sized regions above and below the extracted profile. These regions are individually summed and then averaged together before being subtracted from the 1D Ly$\alpha$ spectrum. This is an imperfect method given the presence of small emission features within the selected background regions, but it is the only process capable of removing the effects of instrument and cross disperser scatter from the echelle scatter signal. Therefore, we expect that our final scatter measurements are likely lower limits on the true value due to the larger contribution from these emission features.

\begin{figure}
\centering
\includegraphics[width=\textwidth]{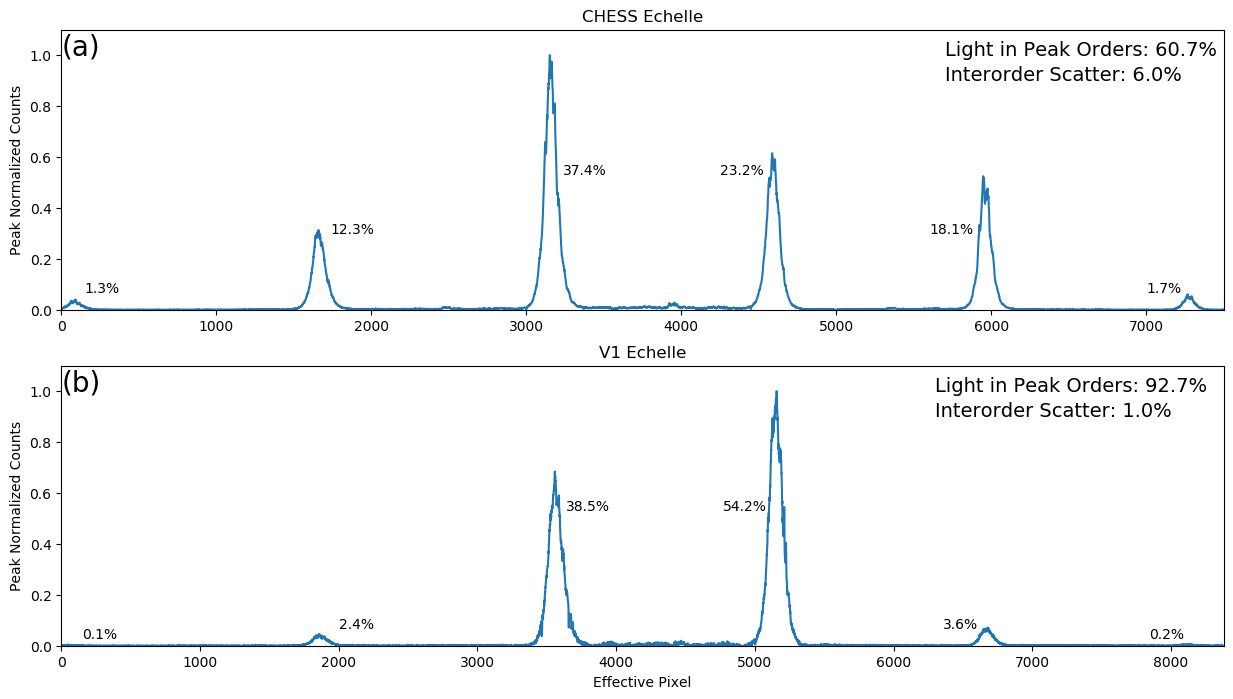}
\caption{The extracted Ly$\alpha$ diffraction profiles from Fig.~\ref{ininst} for the mechanically ruled (a) and V1 (b) echelles. The percent of light in each order is shown next to the corresponding peak. This value is relative to the sum of the light in the entire profile. An estimated interorder scatter is also provided in the top right, measured using the same metric.}
\label{1dinst}
\end{figure}

The two 1D Ly$\alpha$ spectra are shown in Fig.~\ref{1dinst}. While the in-instrument echelle diffraction profile has useful parallels to the echelle profile measured in the square tank, we caution against directly comparing specific order efficiencies between the two given the potential differences in echelle orientation and the unaccounted for contribution of the cross disperser on the observed 
profile. When comparing the spectra of the two gratings, we find that 92.7\% of all of the light captured in the extracted region is confined to the two brightest orders of the V1 grating, compared to 60.7\% for the mechanically ruled grating. This demonstrates how the 50\% increase in peak order efficiency shown in Fig.~\ref{groove_eff} propagates down to the instrument level, specifically, more light is falling on the detector in its flight configuration. We additionally confirm the expected low interorder scatter levels of the V1 grating, finding that $\sim$1\% of the light is scattered between the orders. This is a significant improvement over the 6\% measured for the mechanically ruled echelle, although we note that this value is also much smaller than we had expected at this wavelength. For comparison, Landsman and Bowers measured a peak scatter of 20\% on the STIS E140M echelle grating. This could reflect improvements in modern day mechanical ruling techniques~\cite{stisbad}.

The introduction of a defocus by the Si grating hides further spectral performance enhancements provided by the Si gratings. If the warping was not present, we would additionally expect the Si gratings to produce narrower spectral features in Fig.~\ref{1dinst}. This is equivalent to a higher resolving power brought about by the improved groove positional accuracy and blaze angle consistency that can be achieved using this fabrication process. That would be in addition to the previously described improvements to grating groove efficiency. Future work of this type will be aided by the use of a thicker substrate to prevent the warping of the KOH-etched echelles. We could also gain a more detailed look at how the diffraction effects discussed in Section~\ref{sptstr} impact the instrument.

\subsection{Gains Provided to Flagship Missions} \label{flagship}
The Astro2020 Decadal survey recommended a large UV/visible/IR telescope capable of achieving resolving powers in the FUV of 120,000. Utilizing these new gratings could significantly improve the performance of the spectroscopic channels of this next generation instrument compared to past flagship missions. We can use the LUVOIR/LUMOS-B channel as an example to demonstrate this improvement. This instrument originally had a high resolution channel designed that would utilize gratings similar to those studied in this work~\cite{France19b}. The effective area of LUMOS-B with a conventional grating is $\sim$30,000 cm$^2$ at 1250 \AA~and we assume a STIS-like echelle scatter profile. Observing a faint quasar with F = 1 $\times$ 10$^{-15}$ erg cm$^{-2}$ s$^{-1}$ \AA$^{-1}$ for five hours results in an observation with SNR = 17.2. The introduction of the KOH-etched grating, with its increased efficiency and decreased scatter, could achieve an SNR = 22.9 across the same observing time assuming only the peak orders are included. This demonstrates a potential 33.5\% increase in SNR through the use of Si gratings, which can be further improved upon if more orders can be captured by the detector. This analysis does not include the effects of the write area diffraction pattern discussed in Section~\ref{sptstr}, which would require additional considerations particularly in terms of achievable resolving powers.

\section{Conclusions} \label{conc}
Leveraging a process developed for X-ray diffraction gratings, we have fabricated echelles for the FUV that demonstrate a $\sim$50\% increase in groove efficiency over traditional mechanically ruled echelles (Fig.~\ref{groove_eff}). In-instrument comparisons (Figs.~\ref{ininst} and~\ref{1dinst}) show how the improved peak efficiency propagates down to the instrument level. We additionally found the KOH-etched gratings had an 83\% decrease in interorder scatter compared to the mechanically ruled grating. With the proper instrument design and detector selection, echelle groove efficiencies approaching 80\% with minimal interorder scatter are possible when KOH-etched echelle gratings are used.

To support the continuing development of these gratings, we underwent a detailed accounting of the light budget across the full diffraction arc of the grating. We find that our deviation from the geometrically predicted efficiencies is not due to missing light in the system. We instead find that the remainder of the light resides in neighboring orders distributed around the blaze order. We observed that the individual e-beam write areas introduce a square aperture diffraction effect to each order. We expect that this would not significantly impact the performance of CHESS but could not confirm this due to the substrate curvature preventing a focused in-instrument spectrum. A multipass e-beam exposure, where each region is progressively patterned multiple times but with a small positional offset, could potentially be utilized to smooth out the write field boundaries and eliminate this effect at the cost of doubling the e-beam write time. The feasibility of this process and alternative solutions is still under study.

Future work on KOH echelles will focus on further driving more power into the blaze order. While the rounding out of the otherwise sharp grating facets by the coating could be driving our observed efficiency distribution, our limited data on the pre- vs post-coating V2 grating demonstrates only a modest impact on the diffraction profile (Fig.~\ref{sa_scans}) but does require further study. Fully diagnosing the observed diffraction arc of KOH-etched echelle gratings will require a combination of modeling, fabrication, and characterization efforts. Previous simulations of echelle gratings demonstrate that achieving a perfectly blazed diffraction profile can be complicated, indicating that our geometric models are likely too simplified~\cite{Loewen77, Loewen95}. Additional efficiency gains can be realized through minimizing or removing the Si plateaus. One option to circumvent the plateaus is to replicate the grating, see~\cite{Miles18} for complete details. This process has the benefit of removing the plateaus but does impact the blaze profile, which must be accounted for in the grating design.

Beyond echelles and high resolution spectroscopy, anisotropically etched Si gratings hold potential for a variety of UV gratings. The atomically smooth facets should be comparable to the performance of holographically ruled gratings while simultaneously achieving a blazed profile without requiring post-processing. The blaze angle can be customized, including angles below what is typically achievable for holographic gratings. The groove trace is also free form, so holographic solutions can be reproduced on Si with the added benefit of reducing the restriction on available recording wavelengths, widening the recording geometry parameter space. To support the fabrication of such gratings, we have developed a methodology for translating holographic solutions into an e-beam readable format, which was demonstrated on an EUV grating sample as a part of the ESCAPE Phase A study~\cite{Kruczek21,Grise21}.

We plan on leveraging these benefits through the fabrication of several gratings of increasing complexity, ranging from a spherical grating with a uniform line density to an ellipsoid with variable groove spacing~\cite{Erickson19, Fleming16, McCandliss16, Carlson21}. As in this work, these gratings will be modeled after ones that have flown on sounding rocket missions to provide a baseline point of comparison. For the sounding rocket payloads that are still active, we additionally plan to flight test the fabricated gratings. Combined, this work is a crucial step in our ability to meet the efficiency goals of future large UV observatories.

\section*{Acknowledgements}
Part of this work was carried out in the PSU Materials Research Institute and PSU Nanofabrication Lab and the authors would like to thank the staff that supported the grating development efforts that were undertaken there. NK would like to thank Nicholas Nell and Patrick Behr for their support in CHESS alignment and operation. This work was supported by NASA grants NNX16AG28G, 80NSSC19K0450, and 80GSFC20C0087 to the University of Colorado. 

\section*{Disclosures}
Randall McEntaffer and Drew M. Miles have a financial interest in RFD Optics, a company which could potentially benefit from the results of this research. This interest has been reviewed by the University in accordance with its Individual Conflict of Interest policy, for the purpose of maintaining the objectivity and the integrity of research at The Pennsylvania State University.

\section*{Data Availability}
Data underlying the results presented in this paper are not publicly available at this time but may be obtained from the authors upon reasonable request.

%%%%%%%%%% If using BibTeX:
\bibliography{mybib}

\end{document}